\documentclass[]{emulateapj}
\usepackage{color}
\usepackage{natbib}

\newcommand{\mass}{\mathcal{M}}

\shorttitle{I. The Systemic Console}
\shortauthors{Meschiari et al.}

\begin{document}

\title{Systemic: A Testbed for Characterizing the Detection of Extrasolar Planets. I. The Systemic Console Package.}

\author{Stefano Meschiari \altaffilmark{1}, Aaron S. Wolf \altaffilmark{2},  Eugenio Rivera \altaffilmark{1}, Gregory Laughlin \altaffilmark{1} , Steve Vogt \altaffilmark{1}, Paul Butler \altaffilmark{3}}

\altaffiltext{1}{UCO/Lick Observatory, 
Department of Astronomy and Astrophysics, 
University of California at Santa Cruz,
Santa Cruz, CA 95064}
\altaffiltext{2}{Department of Geophysical and Planetary Sciences, Caltech,
    Pasadena, CA 91125}
\altaffiltext{3}{Department of Terrestrial Magnetism, Carnegie Institute of Washington, Washington, DC 20015}

\begin{abstract}
We present the \textit{systemic Console}, a new all-in-one, general-purpose
software package for the analysis and combined multiparameter fitting of Doppler radial 
velocity (RV) and transit timing observations. We give an overview of the computational algorithms implemented
in the Console, and describe the tools offered for streamlining the 
characterization of planetary systems. We illustrate the capabilities of the package
by analyzing an updated radial velocity data set for the HD128311 planetary system. HD128311 harbors a pair of planets that appear to be participating in a 2:1 mean motion resonance. We show that the dynamical configuration cannot be fully determined from the current data. We find that if a planetary system like HD128311 is found to undergo transits, then self-consistent Newtonian fits to combined radial velocity data and a small number of timing measurements of transit midpoints can provide an immediate and vastly improved characterization of the planet's dynamical state. 
\end{abstract}


\section{Introduction}

During the past decade, the characterization of extrasolar planets has become
a major branch of Astronomy. The field is energized by a variety of ground
and space-based detection programs that are meeting with increasing success.
In the past year, the census of extrasolar planets has exceeded 300, and planets
have now been successfully detected using a variety of techniques, including
doppler radial velocity \citep[e.g.][]{MayorQueloz95, Udry07}, transit
photometry \citep[e.g.][]{Henry00, Charbonneau00, Charbonneau07}, microlensing \citep{Bennett09}, astrometry
\citep{Benedict02, Bean09}, 
stellar pulsations \citep{Silvotti07} and even direct imaging \citep{Chauvin05, Kalas08, Marois08}.

The radial velocity method has been used to discover more than
75\% of the known planets,
and continues to be a dominant technique, both in terms of its continued
productivity \citep[e.g.][]{Fischer05} and its ability to accurately probe
planetary architectures into the vicinity of the terrestrial mass region
\citep[e.g][]{Rivera05, Lovis06, Udry07_HARPS, Mayor08}. A number of planets that
were initially detected using radial velocity (e.g. HD 209458b, HD 189733b, 
HD 149026b, Gl 436b, HD17156b and HD80606b) have been later shown to transit as a result of follow-up
photometry, and because the parent stars of these planets are bright, follow-up
characterizations with a variety of methods have been extremely valuable
\citep[e.g.][]{Deming05}.

The planets that have been detected with the radial velocity technique 
comprise a complicated and non-uniform sample. Some systems such as Upsilon
Andromedae \citep{Butler99, Butler06}, GJ 876 
\citep{Marcy98, Marcy01, Rivera05} and HD69830 \citep{Lovis06} have had multiple planets
subject to very accurate orbital characterization within uniform, well-sampled
data sets. Other systems, for example Epsilon Eridani \citep{Benedict06}, draw their support from a variety of observational sources and in some cases have orbital
parameters that are significantly uncertain. Indeed, it is difficult to 
draw a firm boundary between detections that are secure, and those that
may be subject to serious revision or even elimination.

In addition to the large amount of observational work that has gone
into the detection of extrasolar planets, there is a parallel effort by
theorists to explain the emerging distributions of planets within the 
context of theories of planetary formation and evolution. This work spans a
wide variety of bases, but a unifying principle is that much of it depends
on the raw data being supplied by the catalog of extrasolar planets, and
therein lies a difficulty. Dynamicists have traditionally dealt with
planetary orbital elements that are known to exquisite precision. As far
back as the Eighteenth century, the orbital elements of the solar system
planets were known with an accuracy well in excess of our current
orbital determinations for extrasolar planets. Theoretical interpretations
of the extrasolar planetary data is sometimes made without full account of the 
highly varying signal-to-noise of the datasets that make up the catalog. This
problem is exacerbated by the fact that there exists no continuously up-to-date compendium
of known extrasolar planets in which all of the fits are derived using the same
toolset of routines. The systemic collaboration has been established as an effort to
solve this problem.

The plan for this paper is as follows. In \S \ref{sec:Console}, we describe the systemic
Console. In \S \ref{sec:apps}, we show some sample applications of the tools that are incorporated in the Console,
with a particular emphasis on the planetary system orbiting HD128311 \citep{Vogt05}. We show that our
current radial velocity data set for this system is insufficient for characterizing the resonant relation between the planets, and we demonstrate, using synthetic datasets, how the inclusion of transit timing data (were transits to be detected) would almost immediately eliminate this degeneracy. As another example of the versatility of the code, we describe in Appendix \ref{sec:backend} an automated pipeline (the systemic ``backend'') which runs on top of the same program to create a web application that analyses data sets and aggregates fits.
In \S\ref{sec:conc}, we describe the direction of possible future work with the tools that
we have developed, and conclude.

\section{The systemic Console}\label{sec:Console}
\begin{deluxetable*}{lll}
\tabletypesize{\scriptsize}
\tablecaption{List of tools\label{tab:tools}}
\tablehead{{Name}&{Menu command}&{Description}}
\tablecolumns{3}
\startdata
\textbf{Fitting}&&\\
Levenberg-Marquardt	& Edit $\rightarrow$ Polish		& Multidimensional local optimization (\ref{sec:LM}). \\
Simulated Annealing	& Edit $\rightarrow$ Simulated Annealing		& Multidimensional global optimization (\ref{sec:sa}). \\
De-trend				& Edit $\rightarrow$ Detrend		& Removes linear trends from the radial velocity data.\\
Transit fitting		& Options $\rightarrow$ Transit fitting & Adds transits to the $\chi^2$ statistics.\\
&&\\
\textbf{Periodograms}&&\\
Lomb-Scargle Periodogram	& View $\rightarrow$ Periodogram & Identifies periodicities in the full RV dataset\\
&& and estimates FAPs. \\
Lomb-Scargle Periodogram of residuals 	& View $\rightarrow$ Periodogram	of residuals & Identifies periodicities in the residual RV dataset\\
&& and estimates FAPs. \\
Periodogram of sampling 	& View $\rightarrow$ Periodogram of sampling	& Identifies spurious periodicity peaks associated\\
&& with uneven sampling of the radial velocities. \\
&&\\
\textbf{Uncertainty estimation}&&\\
Bootstrap	& View $\rightarrow$ Bootstrap & Estimates uncertainties using the bootstrap routine; \\
&&  plot and export marginal distributions of orbital \\
&& parameters (\ref{sec:bootstrap}).\\
Markov Chain Monte Carlo	& View $\rightarrow$ Markov Chain Monte Carlo & Estimates uncertainties using the MCMC routine; \\
&& check chain convergence; plot and export marginal\\
&& distributions of orbital parameters (\ref{sec:MCMC}).\\
F-test	& View $\rightarrow$ F-test / F-test significance& \\
&&\\
\textbf{Dynamical analysis}&&\\
Dynamical evolution	& View $\rightarrow$ Orbital evolution and stability & Tracks the fully integrated evolution of the orbital \\
&&elements and the stability of the system.\\
Transits prediction	& View $\rightarrow$ Transits prediction & Calculates the distribution of central transit times\\
&& for a given observational window.
\enddata
\end{deluxetable*}

The {\it systemic Console} is a downloadable software package
\footnote{Freely available at \url{http://www.oklo.org}.} that provides 
an intuitive graphical user interface for the fitting of 
planetary signatures, and an associated suite of dynamical analysis tools (Table \ref{tab:tools}). It can also be used as a specialized, programmable calculator and run scripts in non-interactive mode to access its library of numerical routines.
The program is written in the \textit{Java} programming language for cross-platform portability.

\subsection{Radial Velocities}
The systemic Console allows for a choice between two modeling schemes. For
the majority of the known extrasolar planetary systems,
the planets do not experience significant dynamical 
interactions during the time range spanned by a set of radial velocity
observations. In these cases, the radial velocity variation of the star can be represented as
a sum of $N$ Keplerian orbits, each described by orbital elements
(period $P$, mass $\mass$, eccentricity $e$, mean anomaly $M$, and argument of periastron $\varpi$.)

Summed Keplerians provide an adequate model for nearly all of the planetary
systems that have been discovered to date (Appendix \ref{sec:kep}). Kepler's equation is rapidly
solved using a simple iterative scheme, and hence models can be quickly
evaluated \citep[see e.g.][for a discussion of the current state-of-the-art]{Ford09}. 

There are, however, several
exceptions, notably GJ 876 \citep{Rivera05}, HD202206 \citep{Correia05} and HD60532 \citep{Laskar09}
in which a \emph{self-consistent}, or Newtonian fit is 
required. In these cases, planetary interactions are taken into account in the
fit, and the Console adopts an N-planet model of the system
\begin{equation}
{d^{2}{\bf x_i}\over{dt^{2}}}=-\sum_{j=1}^{N} {G \mass_j({\bf x_i - x_j})
\over{\vert {\bf x_i - x_j}\vert^{3}}} \, ,
\end{equation}
with the integrations carried out using either 4th/5th order Runge-Kutta with
adaptive timestep control or Hermite 4th-order integration \citep{Press, HutMakino95}. When an integrated
model is adopted, a system is defined by the osculating orbital
elements of the planets at the epoch of the first observation expressed in
Jacobi coordinates (see Lee \& Peale 2002). The user also has the option of providing
an integration routine.

Finally, the Console allows the velocity offsets between different data sources to be
additional free parameters; this allows sources with different zero-point offsets 
(e.g. radial velocity surveys using different templates) to be combined in the fitting
procedure.

The Console carries out parameter minimization of the so-called reduced
chi-square statistic
\begin{equation}
\chi^{2}_{RV}=\frac{1}{N_{RV}-M_{fit}}\ \sum_{i=1}^{N} \left[\frac{v_i-v(x_i\, ;\, a_1 \ldots a_M)}{\sigma_i}\right]
^{2}\label{eqn:chi-sq}
\end{equation}
of a fit; in the above expression, $N$
is the number of radial velocity data points, and $M_{fit}$ is the number of 
activated parameters, $a_1 \ldots a_M$. As a rule of thumb, a reduced
Chi-square value near unity is indicative of a ``good" fit to the data, suggesting
that the model is a reasonable explanation of the data within the observational
errors. Typically, larger values usually signal an insufficient modeling of the 
data, whereas smaller values imply that the data has been over-fit. However,
this rule is not exact, and should hence be applied with caution. 

\subsection{Transits}\label{sec:transits}
A rapidly growing number of planets (58 as of writing) with a favorably inclined orbital plane are being further characterized with transit timing data \footnote{Gary, B. 2009; \url{http://brucegary.net/AXA/x.html}, accessed 13 March 2009}. Transits enable direct estimations of planetary masses, radii and mean densities, together with period and phase of the transiting planet \citep{Charbonneau07}. Considerable current interest is focused on detection of transit timing variations (TTVs) which can point to the presence of additional perturbing bodies in a given system.

When supplied to the Console, transits data (central primary and
secondary transits timing) is included with the RV data in the following
way. The Console searches for the best-fit orbital parameters by
minimizing over the joint $\chi^2$ statistic 
\begin{equation}
\chi^{2}= \frac{1}{N_{RV}+N_{tr}-M_{fit}}\left[(N_{RV}-M_{fit})\chi^2_{RV} + \chi^2_{tr}\right]
\end{equation}
where $\chi^2_{RV}$ represents the goodness-of-fit for the radial
velocity component of the model, as described above, and $\chi^2_{tr}$
is representative of the transit component. Ideally,
one would fit together all of the radial velocity and transit
photometry data with a single model to jointly invert for the
parameters that describe all available data. In the future, these
capabilities will be incorporated into the Console. Much progress can
still be made, however, by restricting our analysis to observed times
of central transit with error-bars obtained from separate light curve
analyses. These transit time data can then act as separate constraints
on the observed behavior of the system. To ease implementation, we
compare the predicted and observed {\it location} of the planet at the
observed time of central transit, rather than comparing transit
times. Since the orbital velocities are not changing significantly
with respect to the duration of the eclipse, the difference between these approaches is
negligible. We thus use the following equation to define the
goodness-of-fit statistic for the transit component of the model:
\begin{equation}
\chi^2_{tr} = \sum_{i=1}^{N} \left[\frac{\delta x_i}{\sigma_{\delta x, i}}\right]^{2}
\end{equation}
where $\delta x_i$ is the predicted separation perpendicular
to the line of sight at the observed central transits $t_i$, such that 
\begin{equation}
\delta x_i = |x_*(t_i) - x_P (t_i)|, \ \ \ i = 1 .. N
\end{equation}
The error on $\delta x_i$ is estimated from the error on $t_i$ as 
$\sigma_{\delta x, i} = v_{x, P} \sigma_{t_i}$. While we do not
explore it here, it is important to recognize that regularization of
the fit may be warranted in this type of analysis \citep{Press}.
\footnote{ Regularization is a formal statistical method of compromising
between two distinct sources of information. This is
accomplished by adding a relative weighting factor $\lambda$ in front
of one of the components of the overall $\chi^2$ metric, where
the value of $\lambda$ determines the relative importance of the two
components of the goodness-of-fit. There are many different methods
that can be used to choose an appropriate value for the weighting
factor. In this work, we have implicitly chosen the value $\lambda =
1$, corresponding to an equal weighting.}

Since it is routinely possible to achieve small error bars on the central primary transits (100s for ground-based observations down to 10s for HST observations), a best fit found by the Console that includes transit timing may yield extremely precise determinations of the period and mean anomaly at epoch of the transiting planet \citep[e.g.][]{Wittenmyer05, Bean08}.

Detection of central secondary eclipses \citep{Deming07} also places tight bounds on the eccentricity and argument of periastron of the planet. This additional constraint can break degeneracies present when RVs alone are used; for instance, it can discriminate between eccentric single-planet systems and two-planet systems in a 2:1 resonance with circular orbits \citep{AngladaEscude08}. 

Further afield, it can be possible to measure transit timing variations (TTV) in a dynamically interacting planetary configuration and infer the orbital elements of a perturbing, non-transiting body \citep{Holman05, SteffenAgol05, SteffenAgol07}. 

\subsection{Best-fit model estimation}\label{sec:bestfit}

\subsubsection{Periodograms and False Alarm Probabilities}
The Lomb-Scargle (LS) periodogram is an algorithm for time series analysis of unevenly spaced data \citep{Scargle82, HorneBaliunas86, Press}. The LS periodogram is useful for rapidly identifying periodic signals in the observed data, and to residuals to a given fit, without having to fit for the other orbital parameters. The formula for an error-weighted periodogram $P_x(\omega)$  as implemented in the Console is given in \citet{Gilliland87}; the individual weights are taken to be $w_j = 1/\sigma_j^2$. 

An advantage of this method is that its statistical properties are well known and are conducive to the definition of an analytic  \textit{false alarm probability} (FAP) associated with each periodic signal. When the periodogram is normalized by the total variance $p_0(\omega) = P_x(\omega) / \sigma^2$, the estimated probability that a peak as high or higher would occur by chance is given by $\mathrm{Pr}(p_0, N_f) = 1 - [1 - \exp(-p_0)]^{N_f}$, where $N_f$ is the effective number of frequencies. 

Finally, since the unequal spacing of the data can be a source of spurious periodicities (e.g. those associated with the synodic lunar month or yearly observational schedules), the Console also supports plotting of the power spectral window \citep{Deeming75} overlaid over the standard (non-error weighted) periodogram.

\subsubsection{Levenberg-Marquardt (local minimization)}\label{sec:LM}
Given the observations and associated errors, the goal is to obtain a model configuration $\mathbf{y}_{bf}$ (a $5N$ vector of orbital parameters) such that $\chi^2(\mathbf{y}_{bf}) = \min_y \chi^2$; this is usually reported as the ``best-fit'' solution. Typically, the Lomb-Scargle periodogram is used to comb through periodicities in the data; periodicities are removed in order of decreasing half-amplitude $K$ and optimized using line-minimization. This procedure leads to a set of orbital parameters $\mathbf{y}_0$ which is a rough approximation to the best-fit solution, and can be improved with simultaneous multiparameter minimization. For a discussion of the intricacies of the Keplerian fitting process, see \citet{Cumming08}.

Multidimensional parameter minimization can be carried out using the Levenberg-Marquardt algorithm (LM; \citealt{Press}). Given the initial guess $\mathbf{y}_0$, the LM algorithm can quickly converge to a local minimum $\mathbf{y}'$. Good convergence of the LM algorithm is conditional on the choice of the initial guess and a favorable geometry of the $\chi^2(\mathbf{y})$ surface: in particular, the algorithm is sensitive to rugged $\chi^2$ surfaces and can be prone to converging to non-optimal minima. 

\subsubsection{Simulated Annealing (global minimization)}\label{sec:sa}
So-called ``global'' minimization techniques attempt to avoid getting trapped in local minima by adding a degree of randomness at each iteration step, although at a much greater computational cost. Simulated annealing (SA; \citealt{Press}), by analogy to several thermodynamic processes in nature, defines an ``energy'' $E$ as the objective function to minimize and allows for temperature fluctuations between states at different energies as dictated by the current temperature $T_n$; the temperature $T_n$ is lowered with a (problem-dependent) scheduler. This algorithm is particularly appropriate for rugged $\chi^2$ surfaces, or when the initial guess is sufficiently distant from the best-fit solution.

In our problem, the objective function is clearly $\chi^2(\mathbf{y})$. Given a state $\mathbf{y}_n$, the algorithm selects a new configuration $\mathbf{y}_{n+1}$; the new configuration is accepted and kept with a probability $P(n \rightarrow n+1) \sim \exp(- \Delta E/T_n)$ if $E_{n+1} > E_n$, and is always accepted if $E_{n+1} < E_n$ (a downhill step). The temperature is subsequently updated according to the input scheduler, and the process is repeated until a target number of steps $N$ is reached. The fact that uphill steps are \textit{sometimes} accepted (according to the current temperature) lets the algorithm explore a larger portion of the parameter space and makes it less likely to get stuck in a narrow local minimum.  The trial configuration $\mathbf{y}_n{+1}$ is selected using a \emph{proposal distribution}, which is an easy-to-evaluate generator of trial configurations that picks a new set of parameters given the current set of parameters. The default function is a multivariate Gaussian distribution centered on the current step $\mathbf{y}_n$; the variance $\beta_\mu$ can be chosen independently for each parameter. 

The algorithm requires that the following are configured from the user: 
\begin{enumerate}
\item \textit{temperature scheduler:} the default scheduler decreases $T$ according to $T_n = T_0 (1 - n/N)^\alpha$, where $T_0$ and $\alpha$ are input parameters that dictate the initial temperature and cooling rate. The optimal values of $T_0$ and $\alpha$ are problem-dependent and quite often may determine whether the routine successfully recovers the true global minimum.
\item \textit{generator of trial configurations:} the default generator is a Gaussian function centered around the current configuration, with the scale parameter vector $\mathbf{\beta}_\mu$ given by the user (an initial value is suggested).
\end{enumerate}

Since the correct recovery of $\mathbf{y}_{bf}$ depends on appropriate choices of $T_0, \alpha, N$ and $\beta_\mu$ that are not known a-priori, the Console allows several SA jobs to run in parallel, improving the chance of convergence to the best-fit model. Reconfigurations, in the form of occasionally jump-starting the routine with the best-ever solution, can also be beneficial to the success of the algorithm.

Other global minimization schemes, such as Genetic Algorithms \citep[e.g.][]{Charbonneau95, Laughlin01}, are being considered for inclusion in the Console's built-in array of tools. They can be easily implemented by the user using the routine library offered by the Console.

Finally, we note that certain planetary systems such as HD80606 \citep{Laughlin09, Gillon09, Pont09} include both photometric and spectroscopic data, and contain planets with high orbital eccentricities. In these cases, the connection between observable quantities and the orbital and physical parameters is highly nonlinear, and a modeling framework that relies purely on $\chi^2$ minimization may have a difficult time recovering the correct system configuration. Future releases of the console will therefore incorporate the option of using a fully Bayesian approach to the fitting problem.

\subsection{Error estimation}
Radial-velocity searches are constantly pushing the envelope towards lower and lower masses, frequently at the threshold of detection, with low signal-to-noise ratios. For this reason, once the best-fit parameters have been identified, it is vital to rigorously characterize their uncertainty. The Console offers two independent methods for estimating uncertainty: synthetic datasets refitting (\textit{bootstrap}) and Markov Chain Monte Carlo (MCMC).

\subsubsection{Bootstrap}\label{sec:bootstrap}
The bootstrap procedure consists of drawing with replacement from the observed data points (RV and central transits) and creating a number of synthetic data sets $A^{S}_{i = 1..N}$. The Levenberg-Marquardt fitting procedure is then applied to each dataset, using the best-fit solution for the real dataset as the initial guess. The distribution of the obtained fitted parameters $\mathbf{y}^{S}_{i = 1..N}$ yield an estimated $\mathbf{\sigma}$ for the scatter of the orbital elements around the true intrinsic orbital parameters. 

The bootstrap algorithm is well known \citep{Press} and in common use for estimating planetary elements uncertainties, but presents a number of disadvantages; namely, that it drives a local minimization routine (and is thus subject to the same pitfalls), and that it has a large computational burden. To partially improve on the first weakness, bootstrap can optionally be preceded by a burn-in phase. The burn-in phase obtains a rough estimate of the scatter in the parameters by running a short bootstrap phase. The error estimate is then used in the actual bootstrap run to perturb the best fit a set number of times (e.g. 10 times) per each synthetic dataset fitting; only the best-fitting final configuration is retained. This helps improving the reliability of the bootstrap routine in some cases.

\subsubsection{Markov Chain Monte Carlo}\label{sec:MCMC}
Markov Chain Monte Carlo \citep[see, e.g.,][for exoplanet related implementations]{Ford05, Gregory05} is an alternative method for estimating uncertainties that does not rely on minimization schemes. The MCMC method generates a sequence (\textit{chain}) of configurations $\mathbf{y}_i$ that is sampled from the (unknown) probability distribution $f(\mathbf{y})$. 
The transition probability between two subsequent configurations $\mathbf{y}_n$ and $\mathbf{y}_{n+1}$  is
\begin{equation}
\alpha(\mathbf{y}_{n+1}|\mathbf{y}_n) = \min\left(\exp\left[\frac{\chi'^2_n - \chi'^2_{n+1}}{2}\right], 1\right)
\end{equation}
Assuming that the observational errors are accurately estimated and approximately Gaussian, this transition function assures that, after discarding an initial burn-in phase, the distribution of generated configurations will be sampled from the unknown probability distribution $f$. 

The algorithm consists of looping over the following steps, given an initial state $\mathbf{y}_0$:
\begin{enumerate}
\item given a state $\mathbf{y}_n$ and a Gaussian generator of trial states with scale parameters $\beta_\mu$ (see \ref{sec:sa}), generate a trial state $\mathbf{y}'$;
\item accept the trial state $\mathbf{y}'$ with a probability $\alpha(\mathbf{y}'|\mathbf{y}_n)$ and set $\mathbf{y}_{n+1} = \mathbf{y}'$, otherwise discard it (downhill steps are again always accepted);
\item set $n = n + 1$;
\end{enumerate}
until some convergence criterion of the chain is satisfied. The MCMC algorithm guarantees convergence to the true distribution $f(\mathbf{y})$, but can explore the parameter space inefficiently depending on the choice of $\beta_\mu$, or may not achieve satisfactory convergence within the chosen $N$ steps. The convergence can be visually monitored by interactive plotting of the marginal distribution of the parameters. The acceptance rate of the MCMC algorithm can be interactively monitored as well; an optimal acceptance rate is $\sim 0.25$ (Gelman et al. 2003). 

As with simulated annealing, multiple MCMC chains can be generated in parallel to provide comparison between the results obtained with different choices of $\beta_\mu$ and chain length, which yield similar results within statistical uncertainties if all chains have converged. More sophisticated Bayesian algorithms, such as parallel tempering MCMC \citep{Gregory05b}, may be implemented by the user by exploiting the programmable interface of the Console.

\section{Applications}\label{sec:apps}
\subsection{Resonance characterization in the HD128311 system}
A high fraction of the detected extrasolar systems with multiple planet are involved in near low-order mean motion resonances (MMRs), with at least four of them (GJ876, HD82943, HD73526 and HD128311) being reported to engage in strong 2:1 resonances. Two planets are in a mean-motion resonance when the periods are in a ratio of small integers, and at least one of the resonant angles librates (i.e. it spans a range smaller than $2\pi$). Resonant angles are linear combinations of $\varpi$ (argument of periastron) and $\lambda = M+\varpi$ (coplanarity is assumed). The relevant resonant angles for a 2:1 resonance are $\theta_1 = 2\lambda_2-\lambda_1-\varpi_2$ and $\Delta\varpi = \varpi_2-\varpi_1$ \citep{MurrayDermott00}.

Radial velocity measurements for HD128311 \citep{Vogt05} indicated that the system is locked in a 2:1 MMR, which ensures the long-term stability of the two giant planets. The best-fitting model was indefinitely stable, with the resonant argument $\theta_1$ librating with an half-amplitude of about 60 degrees; a naive fit using Keplerian ellipses instead of the full N-body model is catastrophically unstable within about 2,000 years. Orbital fits for the systems generated using a Monte Carlo procedure (similar to Section \ref{sec:bootstrap}) yielded a proportion of about 60\% stable systems with $\theta_1$ librating and $\Delta\varpi$ circulating to about 40\% with both arguments librating (\emph{apsidal co-rotation}). The large stellar jitter ($\sim$ 9 m/s) and the relatively long periods of the two planets implies that models with different resonant configuration are equally likely given the radial velocity dataset.

However, whether or not the system is in apsidal co-rotation is a crucial piece of information, since it can provide fundamental clues to the migration and interaction history of the system. Scenarios of slow migration and resonant capture into a 2:1 MMR \citep[e.g.][]{Nelson02, Lee02, Beauge06} consistently result in systems that are librating in both resonant arguments. \citet{Sandor06}, analyzing the specific case of HD128311, showed that after adiabatic migration and capture into MMR, the two planets are in apsidal co-rotation and have very small libration amplitudes. If a definitive prevalence of model fits \emph{not} in apsidal co-rotation were ascertained, then the discrepancy has to be explained in terms of subsequent perturbative events (such as sudden termination of migration or planet-planet scattering) that happen after an adiabatic migration process. Analogous studies have been conducted for GJ876 \citep{Kley05} and HD73526 \citep{Sandor07}.

It is therefore important to distinguish between the two resonant configurations (ideally, at the 90\% confidence level or better); this requires a more precise determination of the orbital parameters, which might be achieved, for instance, with additional RV measurements. 

For this purpose, we present a set of additional Doppler measurements taken between June 2005 and May 2008 using the HIRES spectrometer \citep{Vogt94}. Doppler measurements are taken using the standard iodine cell technique \citep[see][for more details]{Vogt09}. Table \ref{tab:data} lists the updated Keck dataset, giving the time of each observation, the radial velocity and the internal uncertainties.

\begin{deluxetable}{ccc}
\tablecaption{New Radial Velocities for HD128311 (\textit{Sample: full table in electronic version})
\label{tab:data}}
\tablecolumns{3}
\tablehead{{JD}&{RV [m/s]}&{Uncertainty [m/s]}}
\startdata
2450983.82690    & -12.95 &  1.45\\ 
2451200.13787    & -21.49  & 1.92\\
2451342.85836    & 62.75  & 2.05\\ 
2451370.82904    & 105.66  & 1.88\\ 
2451409.74660    & 125.71  & 1.62\\ 
2451410.74909    & 118.14  & 2.01\\ 
2451552.16457    & 68.78  & 1.85\\ 
2451581.17009    & 13.35  & 1.64\\
2451680.02544    & -60.10  & 2.17\\
2451974.16142    & 62.03  & 1.74\\ 
2451982.15276    & 32.30  & 1.48\\ 
2452003.02274    & 12.76  & 1.87\\ 
2452003.90155    & 29.10  & 1.98\\ 
2452005.13013    & 27.90  & 1.55\\ 
2452061.87832    & -40.29  & 1.54\\ 
2452062.86745    & -11.26  & 1.67\\ 
\enddata
\end{deluxetable}

\begin{deluxetable}{rccc}
\tablecaption{Orbital fit parameters\label{tab:fits}}
\tablecolumns{4}
\tablehead{{}&{Fit A}&{Fit B}&{Fit C}}
\startdata
Period (d)			& 466.6 [7.5]		& 469.1 [3.3]		& 464.84 \\
			& 909.5 [21.0]	& 893.5 [6.2]		& 901.63 \\
Mass (M$_J$) 		& 1.59 [0.22]		& 1.79 [0.17]		& 1.72\\
	   		& 3.19 [0.11]		& 3.19 [0.08] 	& 3.13\\
Mean anomaly (deg) 		& 270.6 [31.9]	& 282.2 [16.8]	& 263.10\\
			& 192.0 [23.3]	& 190.0 [13.7]	& 193.33\\
Eccentricity			& 0.36 [0.07]		& 0.33 [0.05]		&0.32\\
			& 0.20 [0.09]		& 0.23 [0.05]		& 0.20\\
Long. of periastron (deg)		& 73.8 [24.8]		& 58.98 [19.6]		& 78.04\\
			& 11.7 [20.0]		& 4.54 [14.4]		& 6.59\\
\enddata
\tablecomments{Fit A: integrated best-fit to the \citet{Vogt05} Keck RV data. Fit B: integrated best-fit to the updated RV data reported in this paper and the HET data reported in \citet{Wittenmyer09}. Fit C: orbital elements used to generate the synthetic datasets.  All elements are defined at epoch JD = 2450983.8269. Uncertainties are reported in brackets.}
\end{deluxetable}

\subsection{Best fit}

We update the analysis of \citet{Vogt05} using the tools built in the Console for both the original data and the updated RVs presented in this paper. The Console is well suited to this task, since it can easily derive self-consistent fits (interactively) and do batch Monte-Carlo dynamical analyses on large sets of orbital parameters (non-interactively).

The two prominent periodicities in the \citet{Vogt05} dataset  were found using the integrated Lomb-Scargle periodogram. A self-consistent (Newtonian) best-fit was then derived using the Levenberg-Marquardt minimization routine; one of the built-in N-body integrators (Hermite) was used to derive the radial velocity curve for each choice of orbital parameters. The final best-fit orbital parameters are listed as Fit A (Table \ref{tab:fits}). The uncertainties for each orbital parameter are found using the bootstrap routine on 10,000 synthetic dataset realizations.

\begin{figure}
\plotone{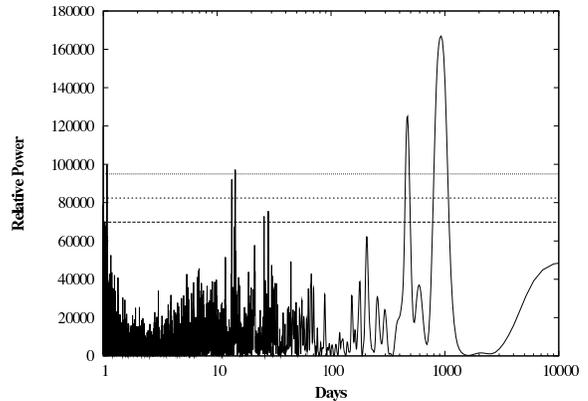}
\caption{Lomb-Scargle periodogram for the combined Keck and HET datasets, as plotted by the Console. Analytical FAPs at levels $10^{-1}$ (long dashed), $10^{-2}$ (short dashed) and $10^{-3}$ (dotted) are overlaid.}\label{fig:power}
\end{figure}

\begin{figure}
\plotone{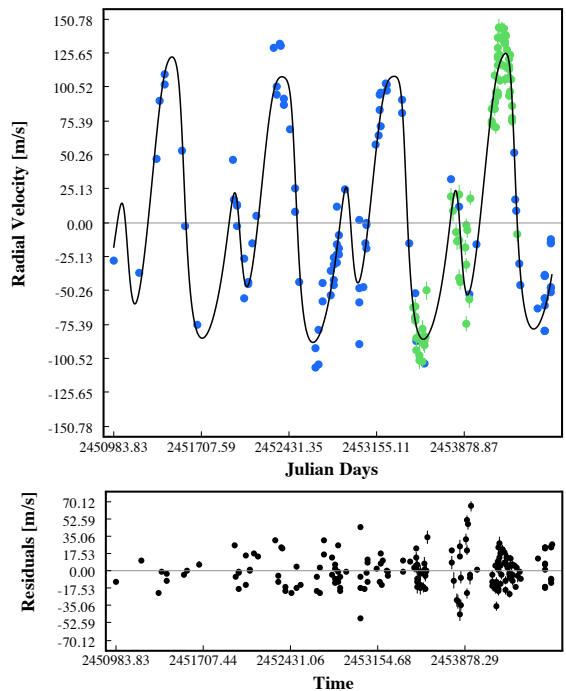}
\caption{Best-fit integrated solution to the RV data presented in this paper (blue) and the HET data (green) reported by Wittenmyer (orbital parameters listed as Fit B in Table \ref{tab:fits}).}\label{fig:het}
\end{figure}

Subsequently, we derived the best-fit for the full updated Keck data (Table \ref{tab:data}), together with the observations taken with the Hobby-Eberly Telescope (HET) and reported in \citet{Wittenmyer09}. The Lomb-Scargle periodogram and the associated analytic FAP estimates are shown in Figure \ref{fig:power}. The Console can account for the zero-point offset and the velocity offset between the two datasets as two additional free parameters. The Newtonian best-fit orbital parameters derived, however, result in a system that is unstable within 1000 years. Therefore, we generated a pool of alternative 5000 bootstrap-generated trial fits, checked each of them for stability within 10000 years and selected the best-fitting stable solution. Its orbital parameters and corresponding uncertainties are listed as Fit B (Table \ref{tab:fits}).  This model is protected by a 2:1 MMR, in which $\theta_1$ librates with amplitude $\sim 60\deg$ and $\Delta\varpi$ circulates. The radial velocity measurements and the star radial velocity curve are shown in Figure \ref{fig:het}.

\begin{figure}

\plotone{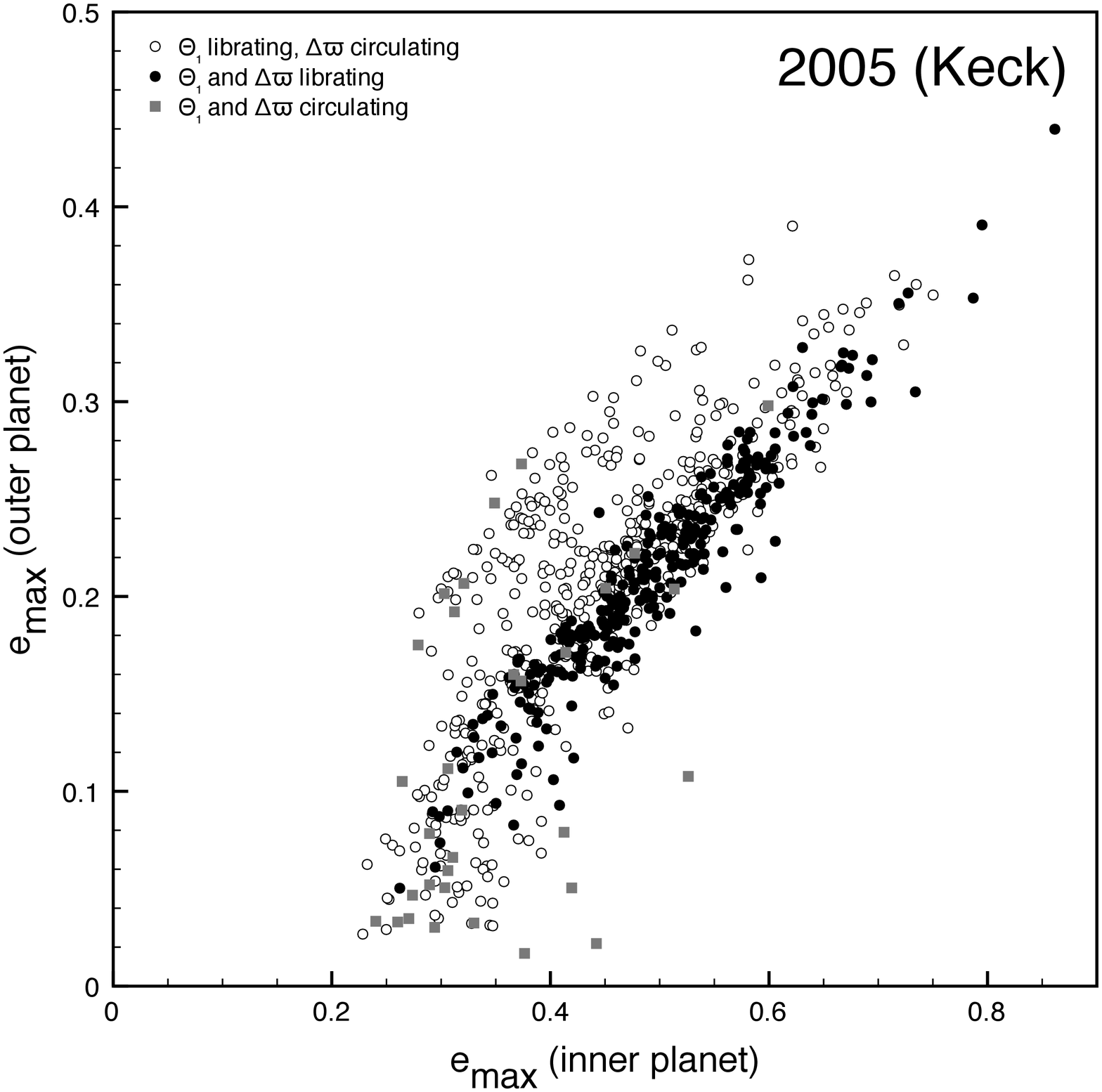}
\plotone{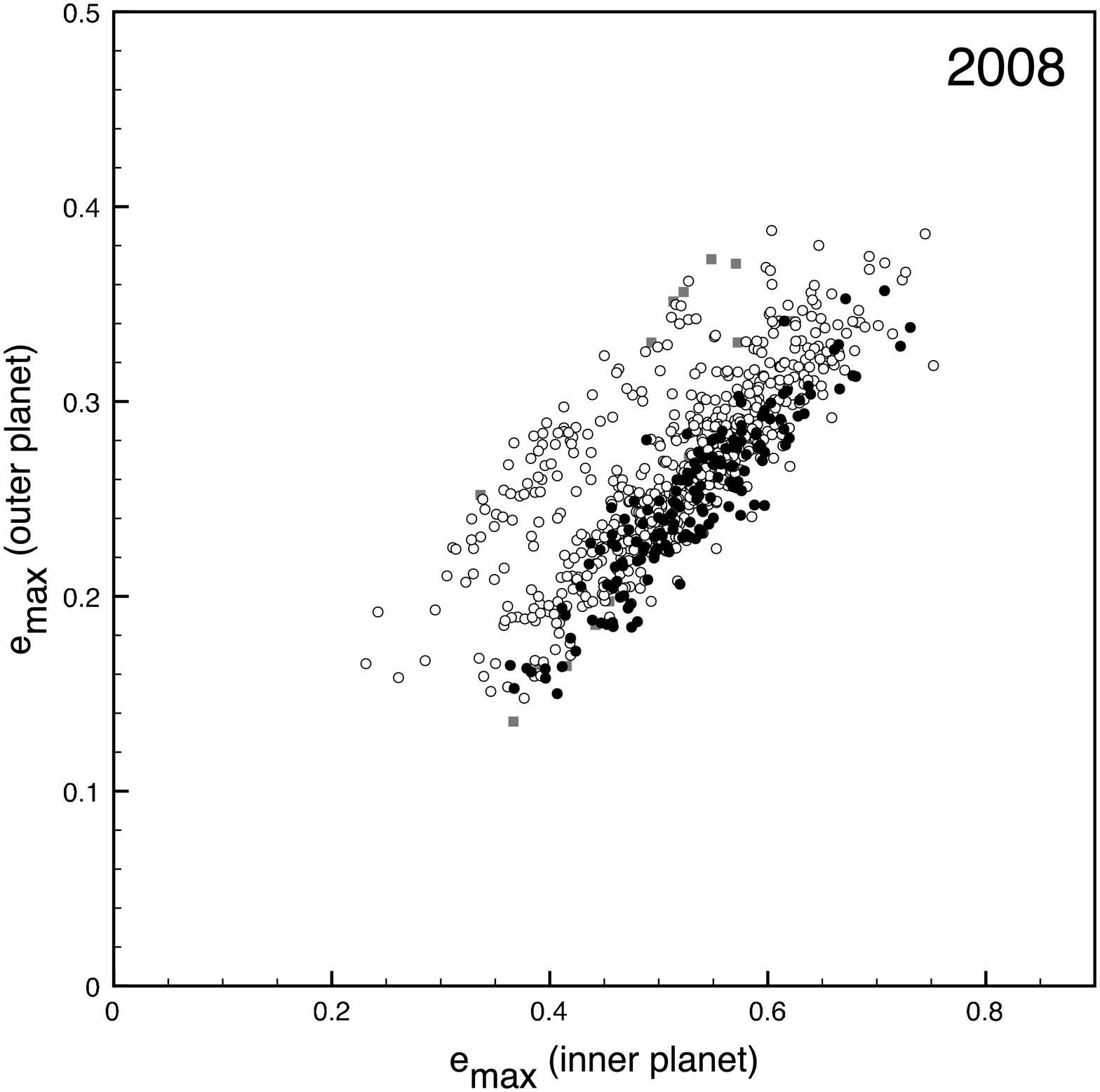}
\plotone{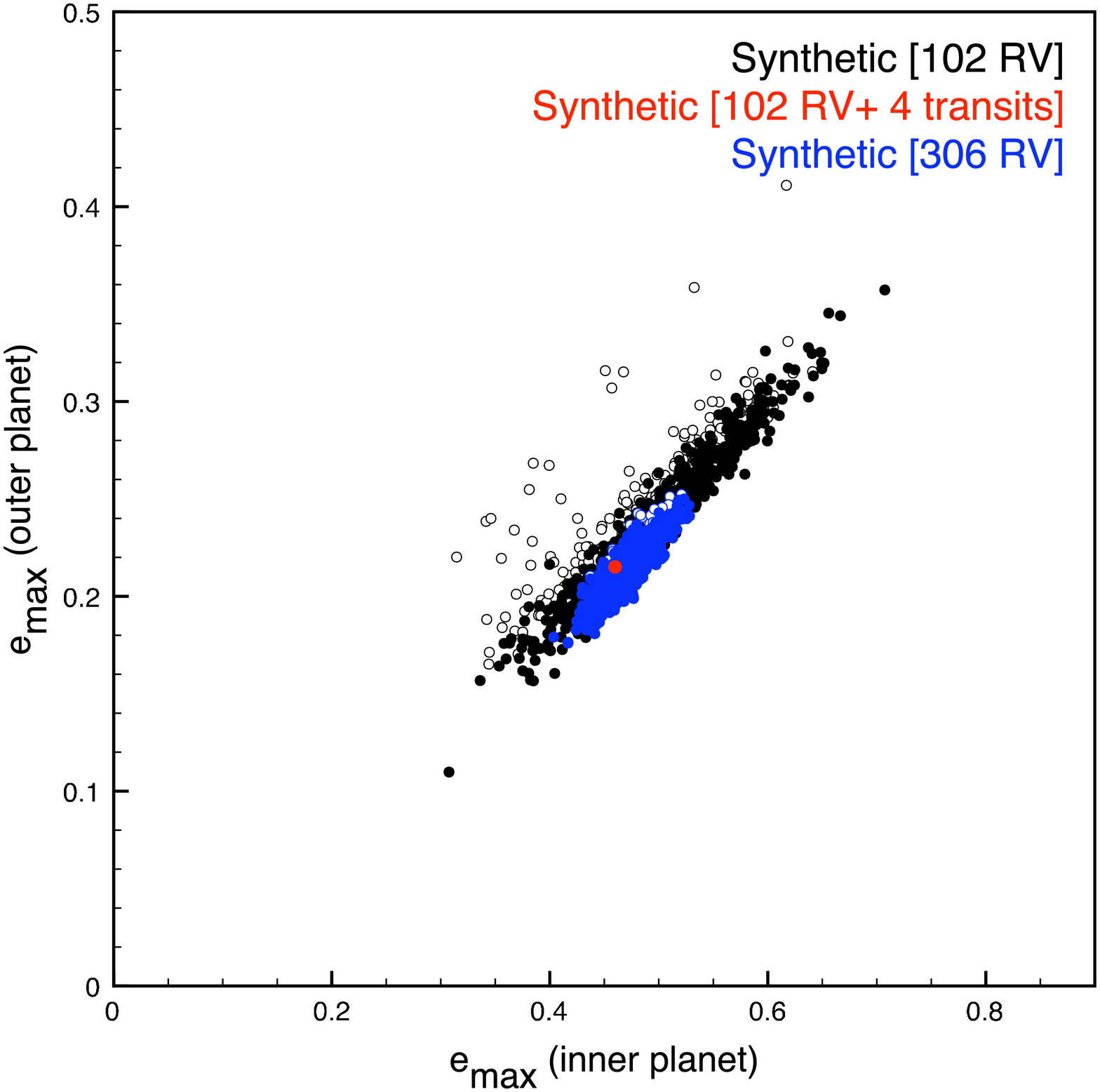}
\caption{Maximum eccentricities observed during $10^4$ yr integrations of self-consistent obtained using the bootstrap routine for data from \citet{Vogt05} (\emph{top}), data presented in this paper (\emph{middle}) and synthetic data (\emph{bottom}). Filled circles: scenarios in which both arguments librate. Open circles: scenarios in which $\theta_1$ librates and $\Delta\varpi$ circulates. Gray squares: scenarios in which both arguments circulate. In the bottom panel, black and blue symbols are for models derived considering RV data only, whereas red symbols are for models considering RV and transits.}\label{fig:hd128311_rv}
\end{figure}

\subsection{Dynamical interactions}

Following the procedure detailed in \citet{Vogt05}, we took the two self-consistent two-planets fits (Fit A and Fit B) and applied a Monte-Carlo bootstrap procedure, in which new fits are derived by resampling with replacements the radial velocity datasets. We created two Monte-Carlo generated libraries of 5000 self-consistent models for two radial velocities datasets: the radial velocities listed in \citet{Vogt05} and the updated Keck data reported in Table \ref{tab:data}. For each of the two groups, 800 fits, stable for at least $10^4$ years\footnote{For longer-term integrations, the builtin integration schemes (RK45 and 4th order Hermite) might not be sufficiently accurate and can be substituted by integration schemes supplied by the user. Alternatively, the Console can be set up to drive packages such as SWIFT (\url{http://www.boulder.swri.edu/$\sim$hal/swift.html}) or Mercury \citep{Chambers97}.}, were selected and integrated forward, recording the maximum eccentricity for both planets and the amplitude of libration of both resonant angles. The results of this analysis are shown in Figure \ref{fig:hd128311_rv}.

With the new radial velocity data, the percentage of model fits that are stable and in apsidal co-rotation using the additional RVs falls slightly to 20\%. A different run considering 1600 models also yields a similar percentage, confirming that the result is robust. The inclusion of the HET data also does not change our result significantly (Table \ref{tab:data}). Therefore, while we have strengthened the case for models of HD128311 that only librate in $\theta_1$, a secure determination of the libration amplitude of $\Delta\varpi$ might be obtained either by a transit monitoring campaign or yet additional RV measurements.
\begin{deluxetable}{rccc}
\tablecaption{Monte-Carlo analysis results\label{tab:mc}}
\tablecolumns{4}
\tablehead{{Data}&{R2}&{R1}&{NR}}
\startdata
2005 (76 Keck RVs)	& 281 [35\%]	& 489 [61\%] 	& 30 \\
2008 (102 Keck RVs) & 160 [20\%]	& 618 [77\%]	& 22 \\
2008 (102 Keck + 78 HET RVs) & 180 [22\%]	& 615 [77\%]	& 5 \\
Fit C, 102 RVs & 603 [75\%] & 197 [25\%] & 0\\
Fit C, 102 RVs + 4 transits & 800 [100\%] & 0 & 0\\
Fit C, 306 RVs & 743 [93\%] & 52 [7\%] & 5 \\
\enddata
\tablecomments{R2: resonant fits with both arguments librating. R1: resonant fits with $\theta_1$ librating and $\Delta\varpi$ circulating. NR: fits have both arguments circulating.}
\end{deluxetable}
\subsection{Constraints by transits}
\noindent Although the a-priori geometric probability for transits $P_{tr}$
\begin{equation}
P_{tr} = 0.0045 \left(\frac{1 \mathrm{AU}}{a}\right)\left(\frac{R_*+R_{pl}}{R_\odot}\right)\left[\frac{1-e\cos(\frac{\pi}{2}-\varpi)}{1-e^2}\right]
\end{equation}
\citep{Bodenheimer03} is very low for HD128311b ($P_{tr} \approx 0.0032$), 
given the high precision that can be achieved by the addition of transits to the $\chi^2$ budget, it is a worthwhile exercise as a proof of concept. Moreover, other resonant systems have higher transiting probabilities; for instance, planets GJ876b and c have a-priori transit probabilities $\sim 1\%$, though the inclination of the system is unfavorable and no transits have been observed \citep{Shankland06}.

We selected the best-fitting solution in apsidal corotation from the ensemble of systems generated by Monte-Carlo bootstrapping of the RVs presented in Table \ref{tab:data} (Fit C). The orbital elements are listed in Table \ref{tab:fits}. Subsequently, we created a synthetic dataset of RVs and transits by integrating forward in time, using the N-body routines offered by the Console. The RVs are generated by sampling the radial velocity curve at the times listed in Table \ref{tab:data}; the tabulated uncertainties and a jitter of 9 m/s are added to the measurement. The central transit times dataset comprises four points, to which we added a Gaussian noise with amplitude $10^{-4}$ d  (comparable to the uncertainties that can be achieved by ground-based transit observations; e.g. \citealt{Alonso08}).

We repeated the analysis detailed in the previous section by bootstrapping exclusively the RV data (Table \ref{tab:mc}); this yields similar ratios, shifted to favor systems in apsidal corotation (similarly to the generating fit). 

As expected, the inclusion of the four central transit times largely reduces the parameter space that can be spanned by Monte-Carlo explorations. The large excursions in $\chi^2$ and the increased ruggedness of the $\chi^2$ space makes the simple bootstrap algorithm, driving a Levenberg-Marquardt scheme, somewhat inefficient in fully exploring the allowed space of orbital parameters (as anticipated in Section \ref{sec:bootstrap}). We therefore used the Markov-Chain Monte Carlo routine supplied with the Console. A long chain of systems ($5\times 10^5$) was generated; the first 50000 systems were discarded and only 1 every 100 systems were retained, to minimize the correlation between subsequent chain elements. 

The tightness of the orbital parameter uncertainties thus generated ($\Delta P_1/P_1 = 2.1\times 10^{-6}$; $\Delta P_2/P_2 = 3\times 10^{-6}$ d; $\Delta M_1/M_1 = 1.4\times 10^{-3}$; $\Delta M_2/M_2 = 4.2\times 10^{-4}$; $\Delta\varpi_1 = 2.4\times 10^{-3}$; $\Delta \varpi_2 = 1.3\times 10^{-3}$) anticipates that the ratio of correctly recovered resonant configuration will be very high. In fact, with the addition of the four primary transits, all of the systems are correctly identified in apsidal corotation (Table \ref{tab:mc}). The maximum eccentricities achieved by the two planets (Figure \ref{fig:hd128311_rv}) are determined within $10^{-3}$. 

As a comparison, we ran the same procedure against 204 additional RVs (a 30-year observation stretch), derived by sampling the integrated stellar radial velocity with the same schedule used for the Keck dataset. This large amount of additional RVs is required to identify the generating planetary system as apsidally corotating with a fraction $>$90\% of models (Table {\ref{tab:mc}).

\section{Discussion}\label{sec:conc}
In this paper, we have described the features of the systemic software package. This software has been written with extensibility, portability and clarity as guiding principles, and is fully adequate for all but the most demanding exoplanet-related analysis tasks. The Console provides a uniform method for analyzing data stemming from a variety of sources (radial velocities surveys and transits) and allows the efficient recovery of the best-fitting stable planetary configuration, even in presence of strong mutual perturbations. It is provided for free to the scientific community.

As an example application, we have analyzed an updated radial velocity dataset for the pair of resonating planets harbored by HD128311. As first noted by \citet{Vogt05}, the orbital solution to this system is degenerate between apsidally corotating and non-apsidally corotating fits; the additional data sets do not break the degeneracy, owing to the large stellar jitter and long orbital periods. We have used an analysis of synthetic data sets to demonstrate that the detection of a transiting extrasolar planet system with planets participating in a low-order mean motion resonance, such as HD128311, would lead to a rapid determination of the libration widths of the resonant arguments and an attendant understanding in how such systems form and evolve. Additionally, our analysis shows that the parameters of non-transiting planets can be very well constrained through transit timing variations in presence of strong mutual interactions. As noted in Section (\ref{sec:transits}), however, a more detailed analysis may be warranted (in particular regarding the issues of fit regularization and full photometry fitting) and will be the object of a follow-up paper. Finally, we showed that breaking the degeneracy at a comparable level with radial velocities would require a prolonged observation campaign, of 30 years or more.

We plan to improve the current feature set of the Console by (1) adding facilities for fully fitting the raw light curve data of a transit detection, (2) implementing more sophisticated routines for best-fit parameter and uncertainty estimation, and (3) allowing non-coplanar, inclined fits. We note that to date, nearly all of the planetary systems that have been detected with the Doppler radial velocity technique can be satisfactorily modeled (to the precision of the observations) using co-planar models with the inclinations assumed to be $90^{\circ}$. The Console's integration routines and internal system representations are fully three-dimensional, however, and so a forthcoming version will enable non-coplanar fits and will accept astrometric measurements \citep[e.g.][]{Bean09}. With the advent of space missions such as SIM Lite and Gaia, there will be numerous opportunities to accurately discern the three-dimensional orbital configurations of many nearby planetary systems \citep{Unwin08}. Finally, signatures of less obvious effects in the spectroscopic and photometric data sets, such as those expected from general relativity \citep{Wu02} or the excitation of tidal modes in the host star \citep{Wu03}, will require more sophisticated modelling to be properly taken into account.

\acknowledgments

We would like to thank the participants in the systemic project for
contributing a large amount of research effort toward the characterization
of extrasolar planets. The results reported in this paper would not have been
possible without their dedicated participation.
We are grateful to Debra Fischer, Eric Ford, Man Hoi Lee, Doug Lin and Peter Jalowiczor for useful discussions, and the anonymous referee for a very thorough evaluation of the paper and several valuable suggestions.

This research has been supported
by the NSF through CAREER Grant AST-0449986, and by the NASA Astrobiology
Institute through Grant NNG04GK19G. The Console software package may be downloaded for free at \url{http://www.oklo.org}.

\appendix
\section{Summed Keplerians model}\label{sec:kep}
When the perturbations between planets are negligible over the observational window, it is appropriate to model the radial velocity curve as a superposition of $N$ Keplerian orbits of fixed orbital elements: 
\begin{equation}
v_r(t) = \sum_{i=1}^{N}K_i[\cos(\upsilon_i+\varpi_i)+e_i\cos\varpi_i] \, ,
\end{equation}
where the radial velocity half-amplitude, $K_i$, of planet $i$ is given by
\begin{equation}
K_i = \left({2\pi G\over{P_i}}\right)^{1/3}{\mass_i \sin i_i 
\over{(\mass_{\star}+\mass_i)^{2/3}}}
\ {1\over{\sqrt{1-e_i^{2}}}} \, ,
\end{equation}
and where the true anomaly, $\upsilon_i$, is related to the eccentric anomaly, $E_i$, via
\begin{equation}
\tan\left[{\upsilon_i\over{2}}\right] = \sqrt{1+e_i\over{1-e_i}}\tan\left[\frac{E_i}{2}\right].
\end{equation}
The eccentric anomaly, $E_i$, in turn, can be expressed in terms of the 
mean anomaly $M_i={2\pi/P_i}(t-T_{\rm peri, i})$ through Kepler's equation
\begin{equation}
M_i = E_i-e_i\sin E_i\, 
\end{equation}
\citep{MurrayDermott00}.

\section{The Systemic Backend}\label{sec:backend}

The {\it systemic backend} is a web application that showcases the power of the Console as an automated engine for data analysis. It consists of a database of catalog information (stellar properties as well as RV and transit measurements) as published in the astronomical literature, and a catalog of model planetary fits for each star. For this purpose, it uses the Console as its main engine to perform a number of automatic data explorations, whereas the user-facing part uses standard ``Web 2.0'' tools (PHP, MySQL, Javascript and \emph{wikis}) to present a coherent overview of the data. A public backend \footnote {The publicly accessible version is available at \url{http://207.111.201.70/php/backend.php}}  is
available as a proof-of-concept to foster collaboration within the broader community of exoplanet researchers and enthusiasts, and to present and maintain the catalog of fits to radial velocity and transit timing data for known planet-bearing stars. Each user has a personal data page and fit catalog, the possibility of commenting on other team member's fits, and can interact with other team members within a private and secure  environment. A more powerful and customizable version is also available on request for use by planet hunter teams, and can be useful to maintain an integrated database of datasets and models in face of the increasing flux of RV and transit data.

The fit catalog is scanned by a number of Console components,
which continually sift through the uploaded fits in non-interactive mode. One component
implements a bootstrap routine to calculate uncertainties on the
orbital parameters of each fit; data from the bootstrap routine is
stored in a database for creating scatter plots. Two other components check for dynamical instability over periods of 1,000 and 10,000 years, with stability
defined by the rough criterion of requiring a smaller than 1\% fractional
change in semi-major axis with respect to the average semi-major axis
observed
during a full N-body integration. This step flags highly unstable planetary systems
that experience ejections or collisions.
Data from the integration is retained for plotting of orbital evolution and for future additional investigations. Dynamical data (orbital parameters, radial velocity data, fit parameters,
stability, integrations, bootstrap results) is then transparently
presented to the user as a set of web pages and can be aggregated and sliced using a web-based query system.

\bibliographystyle{apj}

\end{document}